\documentclass[12pt]{article}
\usepackage{amsfonts}
\usepackage{pstricks}
\usepackage{pst-node}
\usepackage{epsfig}
\usepackage{amsmath,amssymb,latexsym, accents}
\DeclareMathOperator\arctanh{arctanh}

\usepackage{import}

\usepackage{enumerate}

\usepackage{physics}

\usepackage[lmargin=1.0in, rmargin=1.0in, tmargin=1.0in, bmargin=1.0in]{geometry}

\usepackage{cancel}

\usepackage{empheq}

\usepackage{comment}

\usepackage{setspace}
\begin{document}

%--------------------------------------------
% several abbreviations
%\def\ba{\begin{eqnarray}}
%\def\ea{\end{eqnarray}}
%\def\w{\wedge}
%--------------------------------------------

%\setstretch{1.0}

%\begin{titlepage}
\title{\bf Scale invariant Einstein-Cartan theory in three dimensions}
\author{Muzaffer Adak$^{1,2,}$\footnote{madak@pau.edu.tr} , Nese Ozdemir$^{1,}$\footnote{nozdemir@itu.edu.tr} , Ozcan Sert$^{2,}$\footnote{osert@pau.edu.tr} \\
  {\small $^1$Department of Physics Engineering, Faculty of Sciences and Arts,} \\
  {\small  Istanbul Technical University, Istanbul, Turkey} \\
  {\small $^2$Department of Physics, Faculty of Science, Pamukkale University, Denizli, Turkey} \\
 }
 
  \vskip 1cm
\date{10 January 2023}
\maketitle

 \thispagestyle{empty}

\begin{abstract}
 \noindent
We retreat the well-known Einstein-Cartan theory by slightly modifying the covariant derivative of spinor field by investigating double cover of the Lorentz group. We first write the Lagrangian consisting of the Einstein-Hilbert term, Dirac term and a scalar field term in a non-Riemannian spacetime with curvature and torsion. Then by solving the affine connection analytically we reformulate the theory in the Riemannian spacetime in a self-consistent way. Finally we discuss our results and give future perspectives on the subject.  \\

%\vskip 1.0cm
%\bigskip


% 11.15.Kc    (Gauge field theory) Classical and semiclassical techniques
% 04.50.Kd    Modified theories of gravity
% 02.40.Yy    Geometric mechanics
\end{abstract}
%\end{titlepage}

\section{Introduction}

The Einstein equation could be formulated in the language of exterior algebra in the $n$-dimensional spacetime ($n \geq 3$), 
 \begin{align}
     \widetilde{G}_a := \frac{1}{2} \widetilde{R}^b{}_c \wedge *(e_a \wedge e_b \wedge e^c) = \kappa \widetilde{\tau}_a[matter],
 \end{align}
where $e^a$ is the orthonormal coframe (or orthonormal basis 1-form), $\widetilde{R}^a{}_b$ is the Riemann curvature 2-form, $*$ represents Hodge dual map, $\wedge$ is the exterior product, $\kappa$ is a coupling constant, $\widetilde{\tau}_a[matter]$ denotes energy-momentum ($n-1$)-form of matter and $\widetilde{G}_a$ is the Einstein tensor ($n-1$)-form. In four dimensions Einstein tensor 3-form has 16 components. On the other hand, the Riemannian curvature 2-form has 20 independent components (36 from $\widetilde{R}^a{}_{b}$ minus 16 from the Bianchi identity, $\widetilde{R}^a{}_{b} \wedge e^b=0$). Thus in vacuum, $\widetilde{\tau}_a[matter]=0$, though all components of the Einstein tensor vanish, some components of $\widetilde{R}^a{}_b$ may still live and then gravitational waves are allowed in an empty spacetime.

Similarly if one does the same analysis in the three-dimensional spacetime, it is seen that there are 9 components at $\widetilde{G}_a$ and 6 independent components at $\widetilde{R}^a{}_b$. Consequently, as the Einstein tensor vanishes, all components of $\widetilde{R}^a{}_{b}$ must also be zero. It means that there can not be gravitational waves in vacuum. Correspondingly in three-dimensions the bare Einstein's general relativity is not a dynamical theory. Therefore there is a wide literature on modified general relativity in three dimensions \cite{deser_jackiw_temp_1982}-\cite{hakan_tekin_2021}.

One of modifications is to go beyond the Riemannian geometry. Firstly we can enlarge it by allowing torsion. Since it is thought that torsion tensor is sourced by fermionic matter, it is natural to extend it to couple a Dirac spinor to three-dimensional Einstein theory. For that we need to know the Lorentz-covariant exterior derivative of a spinor, $\psi$, and its adjoint, $\overline{\psi}$. They are done by the formulas,
 \begin{align}
     D\psi = d\psi + \frac{1}{2} \omega^{ab}\sigma_{ab} \psi  \qquad \text{and} \qquad  D\overline{\psi} = d\overline{\psi} - \frac{1}{2} \overline{\psi} \sigma_{ab} \omega^{ab}, \label{eq:cov-deriv-spinor}
 \end{align}
where $d$ is the exterior derivative, $\sigma_{ab}=-\sigma_{ba}$ is the generator of the restricted special Lorentz group, $SO_+(1,2)$, and $\omega_{ab}= -\omega_{ba}$ is the connection 1-form for the orthonormal frame bundle. On the other hand, it is known that $SO_+(1,2)$ is doubly covered by $Spin_+(1,2)$ group which is also the four-dimensional even subalgebra, $Cl^+(1,2)$, of the eight-dimensional Clifford algebra, $Cl(1,2)$. Meanwhile, a basis set of $Cl^+(1,2)$ is given by $\{1, \sigma_{ab}\}$. As its $\sigma_{ab}$ element generates the Lorentz transformation via the exponentiation $S = e^{\frac{1}{2}\sigma_{ab}\vartheta^{ab}(x)}$, the unit element generates a scale transformation via $ W = e^{1 f(x)}=e^{f(x)} \in \mathbb{R}^+$ where $\vartheta^{ab}(x)$ and $f(x)$ are the concerned transformation parameters. A Lorentz transformation of any two orthonormal coframes could be written $\gamma' = S \gamma S^{-1}$ in terms of $Cl(1,2)$-valued 1-form $\gamma = \gamma_a e^a$ where $2\sigma_{ab}=\frac{1}{2}(\gamma_a\gamma_b - \gamma_b \gamma_a)$ and $\eta_{ab}=\frac{1}{2}(\gamma_a\gamma_b + \gamma_b \gamma_a)$. Since $Cl^+(1,2)$ is four-dimensional, $\gamma_a$ can be represented by real $2\times 2$ matrices in which case a spinor $\psi$ is represented by a two-component complex column matrix and transforms with respect to $\psi= S \psi$ under a Lorentz transformation represented by $2\times 2$ matrix, $S$. In this work we aim to extend the covariant derivative of a spinor given by the equation (\ref{eq:cov-deriv-spinor}) as to include rescaling generated by $I$ of $Spin_+(1,2)$ group. As a scale transformation gives rise to $e^a \to W e^a$ on the orthonormal coframe, there are two possibilities for the affine connection: $\omega^a{}_b \to \omega^a{}_b$ or $\omega^a{}_b \to \omega^a{}_b - \delta^a_b W^{-1}dW$. Both of them leave the curvature 2-form invariant. Here we adhere the first option, because we want to work a modification of the Einstein-Cartan theory which is formulated in the Riemann-Cartan  spacetime with a metric compatible connection before and after a rescaling. Thus we leave the affine connection scale-invariant like in Ref.\cite{tekin-robin-1982-PLB}-\cite{obukhov-1982-PLA}. Now by combining two transformations we define the Weyl group $W(2,2) := SO_+(1,2)\otimes \mathbb{R}^+$ with four parameters, $ \{ a_{01},a_{02},a_{12}, f \}$. Consequently, we postulate the transformation rules for some basic quantities under a $W(2,2)$-transformation,
  \begin{subequations}
    \begin{align}
     e^{a'} &= W L^{a'}{}_a e^a \quad \text{with} \quad \eta_{a'b'} = L^{a}{}_{a'} L^{b}{}_{b'} \eta_{ab} \quad \text{so} \quad \iota_{a'}= W^{-1} L^a{}_{a'} \iota_a , \\
       \omega^{a'}{}_{b'} &=  L^{a'}{}_a \omega^a{}_b L^b{}_{b'} + L^{a'}{}_a d L^a{}_{b'} , \\
       \psi' &= W^{-1} S \psi \quad \text{and} \quad \overline{\psi'} = W^{-1} \overline{\psi} S^{-1},
     \end{align}
  \end{subequations}
where $\iota_a:= \iota_{X_a}$ denotes interior product with respect the orthonormal base vector $X_a$ ($\iota_b e^a=\delta^a_b$), the transformation elements $L^{a'}{}_a$ and $L^a{}_{a'}$ ($L^a{}_{a'} L^{a'}{}_b=\delta^a_b$) form the Lorentz group $SO_+(1,2)$ which is generated by $\sigma_{ab}$ and the other transformation element $W \in \mathbb{R}^+$ which is generated by $I$. Both generators together are bases of $Cl^+(1,2)$ algebra. Accordingly the transformations of non-metricity, torsion and curvature are calculated readily,
 \begin{subequations}
     \begin{align}
         Q_{a'b'} &=  L^{a}{}_{a'} L^{b}{}_{b'} Q_{ab} , \\
         T^{a'} &= W  L^{a'}{}_{a} \left( T^a + W^{-1} dW \wedge e^a \right),\\
         R^{a'}{}_{b'} &=  L^{a'}{}_{a} R^{a}{}_{b} L^b{}_{b'} .
     \end{align}
 \end{subequations}
It is worthy to notice that non-metricity and curvature are scale-invariant, but torsion is not. Nonetheless, additive contribution in torsion transformation will be useful at extension of covariant derivative of spinor. More specifically we will need $W(2,2)$-transformed trace 1-form of torsion, $T=\iota_a T^a$,
 \begin{align}
     T' = T - 2 W^{-1} dW .
 \end{align}
Then, we write $W(2,2)$-covariant exterior derivative of a spinor, $ \mathbb{D} \psi$, and its adjoint, $\mathbb{D}\overline{\psi} := (\mathbb{D}\psi)^\dagger \gamma_{0}$,
 \begin{align} \label{eq:cov-derits-spinors}
     \mathbb{D}\psi = d\psi + \Omega \psi - \frac{1}{2}IT \psi \qquad \text{and} \qquad       \mathbb{D}\overline{\psi} = d{\overline{\psi}}- \overline{\psi} \Omega - \frac{1}{2}IT \overline{\psi}.
 \end{align}
The term $IT\psi/2$ is novelty of this paper. In our prescription the torsion trace stands for the Weyl gauge. A similar conclusion was remarked in Obukhov's paper \cite{obukhov-1982-PLA}. Here the quantity $ \Omega := \frac{1}{2} \omega^{ab} \sigma_{ab}$ must transform according to
 \begin{align}
     \Omega' = S \Omega S^{-1} + S dS^{-1}
 \end{align}
for $\mathbb{D}\psi$ and $ \mathbb{D}\overline{\psi}$ to transform in covariant way, i.e., $ \mathbb{D}\psi' = W^{-1} S (\mathbb{D}\psi)$ and $\mathbb{D}\overline{\psi'} = W^{-1} (\mathbb{D}\overline{\psi}) S^{-1}$. It is worthwhile to remind that the transformation elements are generated by all bases, $\{I, \sigma_{ab}\}$, of $Cl^+(1,2)$ Clifford algebra as
  \begin{align}
      S = e^{\frac{1}{2}\sigma_{ab}\vartheta^{ab}(x)} \in Spin_+(1,2) \quad \text{and} \quad W = e^{I f(x)} \in \mathbb{R}^+ .
  \end{align}

On the other hand, there is an inconsistency in the formulation of the standard Einstein-Cartan theory that is often overlooked. To see this problem explicitly we remind two formulations of the Dirac theory as being equation approach and Lagrangian approach. Firstly a spinor, $\psi$, and its exterior derivative, $d\psi$, are defined for both formulations. Then the covariant exterior derivative of spinor, $D\psi$, is postulated via the minimal coupling principle meaning simply to replace $d$ with $D$. Finally by following the equation approach the Dirac equation is written by
 \begin{align}
     *\gamma \wedge D\psi + m \psi *1 =0 .
 \end{align}
However, the Dirac equation obtained from the Dirac Lagrangian by an independent variation is 
 \begin{align}
     *\gamma \wedge \left(D - \frac{1}{2} T \right)\psi + m \psi *1 =0 .
 \end{align}
The term, $T/2$, causes an inconsistency at the two formulations. Our novel definition (\ref{eq:cov-derits-spinors}) in this paper will remedy this matter as well. Besides, when we compare the constructions of $SO_+(1,2)$-covariant exterior derivative, $D\psi$, and $W(2,2)$-covariant exterior derivative, $\mathbb{D}\psi$, of spinor, we realize that the latter does not follow the minimal coupling recipe.

In the next section we summarize our notations, conventions and definitions. Then we formulate the extended Einstein-Cartan theory by giving a Lagrangian 3-form which is invariant under a $W(2,2)$-transformation. After obtaining variational equations from that we solve analytically the affine connection in terms of spinor field and scalar field which has to be introduced for the scale invariance of Einstein-Hilbert Lagrangian. By substituting our findings into other field equations we rewrite them Riemannian terms plus new terms coming from torsion of geometry. We especially trace ones which are caused by our novel contribution in the covariant derivative of spinor. At last step, we insert the calculated affine connection back to the total Lagrangian 3-form by adding a constraint term, $\lambda_a \wedge T^a$, for zero-torsion, then we compute field equations by varying this total Riemannian Lagrangian. At the end we observe that both formulations are equivalent, but notice that the non-Riemannian one is tidier. 

\section{Notations, conventions, definitions}

The triple $\{M,g,\omega\}$ defines a metric affine geometry where $M$ is three-dimensional orientable and differentiable manifold, $g$ is symmetric and non-degenerate metric, $\omega$ represents the metric compatible full (or affine) connection \cite{thirring1997}-\cite{lounesto1997}. We denote the orthonormal coframe by $e^a$, then write metric as $g=\eta_{ab} e^a \otimes e^b$ where $\eta_{ab}$ is the Minkowski metric with the signature $(-,+,+)$. In the language of exterior algebra, $e^a$ is called orthonormal 1-form and the Cartan structure equations are given by non-metricity 1-form, torsion 2-form and curvature 2-form tensors, respectively,
 \begin{subequations}\label{eq:cartan-ort}
 \begin{align}
     Q_{ab} &:= -\frac{1}{2} D\eta_{ab} = \frac{1}{2} (\omega_{ab} + \omega_{ba})=0, \label{eq:nonmetric}\\
     T^a &:= De^a = de^a + \omega^a{}_b \wedge e^b \neq 0, \label{eq:tors}\\
     R^a{}_b &:= D\omega^a{}_b := d \omega^a{}_b + \omega^a{}_c \wedge \omega^c{}_b \neq 0. \label{eq:curv}
 \end{align}
 \end{subequations}
The metric compatibility condition (\ref{eq:nonmetric}) yields that the full connection 1-form is anti-symmetric, $\omega_{ab}=-\omega_{ba}$. Accordingly, it can be decomposed uniquely to Riemannian piece, $\widetilde{\omega}_{ab}$ and non-Riemannian piece, $K_{ab}$,
 \begin{align}
     \omega_{ab}=\widetilde{\omega}_{ab} + K_{ab}, \label{eq:connec-decom}
 \end{align}
where $\widetilde{\omega}_{ab} = - \widetilde{\omega}_{ba}$ is the Levi-Civita connection 1-form and  $K_{ab} = - K_{ba}$ is the contortion tensor 1-form
 \begin{subequations}
  \begin{align}
   \widetilde{\omega}_{ab} &= \frac{1}{2} \left[ -\iota_a de_b + \iota_b de_a + (\iota_a \iota_b de_c) e^c \right] &  &\text{or} & \widetilde{\omega}^a{}_b \wedge e^b &= -de^a , \label{eq:Levi-Civita}\\
     K_{ab} &= \frac{1}{2} \left[ \iota_a T_b - \iota_b T_a - (\iota_a \iota_b T_c) e^c \right] &  &\text{or} &  K^a{}_b \wedge e^b &= T^a . \label{eq:contortion}
      \end{align}
 \end{subequations}
By substituting (\ref{eq:connec-decom}) into (\ref{eq:curv}) we decompose the full curvature as well
 \begin{align}
     R^a{}_b = \widetilde{R}^a{}_b + \widetilde{D}K^a{}_b + K^a{}_c \wedge K^c{}_b \label{eq:decomp-curva}
 \end{align}
where  $\widetilde{R}^a{}_b$ is the Riemannian curvature 2-form and $ \widetilde{D} $ denotes the covariant exterior derivative with respect to $\widetilde{\omega}^a{}_b$,
 \begin{subequations}
   \begin{align}
       \widetilde{R}^a{}_b &:= d \widetilde{\omega}^a{}_b + \widetilde{\omega}^a{}_c \wedge \widetilde{\omega}^c{}_b  ,\\ 
        \widetilde{D}K^a{}_b &:= d K^a{}_b + \widetilde{\omega}^a{}_c \wedge K^c{}_b - \widetilde{\omega}^c{}_b \wedge K^a{}_c .
   \end{align}
 \end{subequations}
All the Riemannian quantities will be labelled by a tilde over them in this paper. Some useful notations and identities are listed as
 \begin{subequations} \label{eq:identities1}
 \begin{align}
     e^{ab\cdots} &:= e^a \wedge e^b \wedge \cdots , \qquad \iota_{ab \cdots } := \iota_a \iota_b \cdots , \qquad \iota_b e^a = \delta^a_b, \\
      e^a \wedge *e^b &= \eta^{ab} *1 , \qquad e^a \wedge *e^{bc} = -\eta^{ab} *e^c + \eta^{ac} *e^b, \\
     e^a \iota_a \Theta &= p \Theta , \qquad *(\Theta \wedge e_a) = \iota_a * \Theta, \qquad \Theta \wedge * \Phi = \Phi \wedge *\Theta \\
     *1 &= \frac{1}{3!} \epsilon_{abc} e^{abc} , \qquad *e_a = \frac{1}{2!} \epsilon_{abc} e^{bc} , \qquad *e_{ab} = \epsilon_{abc} e^c , \qquad *e_{abc} = \epsilon_{abc}, \\
     \epsilon^{abl}\epsilon_{abc} &=-2! \delta^l_c, \qquad \epsilon^{akl}\epsilon_{abc}=-\left(   \delta^k_b \delta^l_c - \delta^k_c \delta^l_b \right), \qquad \epsilon^{abc}\epsilon_{klm} =- 
     \begin{vmatrix}
       \delta^a_k & \delta^a_l & \delta^a_m\\
       \delta^b_k & \delta^b_l & \delta^b_m\\
       \delta^c_k & \delta^c_l & \delta^c_m
     \end{vmatrix} , \\
     D*e_a &= *e_{ab} \wedge T^b , \qquad  D*e_{ab} = *e_{abc} \wedge T^c , \qquad  D*e_{abc} = D\epsilon_{abc} =0 ,
 \end{align}
 \end{subequations}
where $\Theta$ and $\Phi$ are some $p$-forms, $\epsilon_{abc}$ is the totally anti-symmetric epsilon symbol with $\epsilon_{012}=+1$ and $\delta^a_b$ is the Kronecker delta\footnote{When $Q_{ab}\neq 0$, it is $D*e_{abc}=D\epsilon_{abc}=-Q \epsilon_{abc}$ where $Q=\eta_{ab}Q^{ab}$.}.

Four-dimensional even subalgebra, $Cl^+(1,2)$ of eight-dimensional Clifford algebra, $Cl(1,2)$, is generated by the unit matrix, $I$ and the gamma matrices $\gamma_a$ satisfying the condition
 \begin{align}
     \gamma_a \gamma_b + \gamma_b \gamma_a = 2 \eta_{ab} I .
 \end{align}
One can consult the Appendix for details about the Clifford algebra. We choose the real representations of the gamma matrices given in the equation (\ref{eq:dirac-matricies}), in which case the basis set of $Cl^+(1,2)$ is given by $\{ I, \sigma_{ab} \}$ or $ \{ \gamma_5, \gamma_{a} \}$ because of the results, 
 \begin{align}
     \gamma_5 := \gamma_0 \gamma_1 \gamma_2 = I , \qquad \sigma_{ab} := \frac{1}{4} [\gamma_a, \gamma_b] = \frac{1}{2} \epsilon_{abc} \gamma^c .
 \end{align}
Consequently we will encounter two independent covariant bilinears  \begin{align}
    \rho := \overline{\psi} \psi \qquad \text{and} \qquad j_a :=  \overline{\psi} \gamma_a \psi , \label{eq:bilinears}
\end{align}
where $\overline{\psi}$ is the Dirac adjoint of $\psi$. As special to three dimension one can write $q_{ab} := \overline{\psi} \sigma_{ab} \psi = \frac{1}{2} \epsilon_{abc}j^c$. Here $q_{ab}$ is a quantity related with particle's electromagnetic moment on which could be made observations in particle physics laboratories. In fact, $q^{ab} \sigma_{ab}$ is the probability density of electromagnetic moment of particle \cite{lounesto1997}. They satisfy $ \rho^\dagger = -\rho$ and $j_a^\dagger = j_a$ and $q_{ab}^\dagger = q_{ab}$. The identities below will be helpful in the calculations,
  \begin{subequations} \label{eq:identities2}
 \begin{align}
     \gamma_a \gamma_b &= \eta_{ab} I + \epsilon_{abc} \gamma^c , \\
     \sigma_{ab} \gamma_c - \gamma_c \sigma_{ab} &= \eta_{bc} \gamma_a - \eta_{ac} \gamma_b , \\
     \sigma_{ab} \gamma_c + \gamma_c \sigma_{ab} &= \epsilon_{abc} I , \\
     [\sigma_{ab} , \sigma_{cd}] &= -\eta_{ac}\sigma_{bd} + \eta_{ad}\sigma_{bc} + \eta_{bc}\sigma_{ad} - \eta_{bd}\sigma_{ac} , \\
     \gamma_0 I^\dagger \gamma_0 &= -I , \qquad \gamma_0 \gamma_a^\dagger \gamma_0 = \gamma_a , \qquad \gamma_0 \sigma_{ab}^\dagger \gamma_0 = \sigma_{ab} , \\
     \gamma_0^\dagger &= -\gamma_0 , \qquad \gamma_1^\dagger = \gamma_1 , \qquad \gamma_2^\dagger = \gamma_2 ,  
 \end{align}
 \end{subequations}
where the symbol ${}^\dagger$ denotes the Hermitian conjugation. In this representation, spinor field $\psi$ can be considered by a two-component complex column matrix and its Dirac adjoint is defined by $\overline{\psi}:= \psi^\dagger \mathcal{C}$ where $\mathcal{C}$ is the charge conjugation matrix satisfying the relation, $ \mathcal{C} \gamma_a \mathcal{C}^{-1} = - \gamma_a^T$. Here ${}^T$ means transpose matrix. As a complementary remark we remind that the charge conjugated spinor is defined by $\psi_{\mathcal{C}}:= \mathcal{C} \overline{\psi}^T$. In our representation we will use $\mathcal{C} = \gamma_0$ meaning explicitly $\overline{\psi}:= \psi^\dagger \gamma_0$. Correspondingly, after discussions done in Introduction we wrote the $W(2,2)$-covariant exterior derivative of $\psi$ and $\overline{\psi}$ by (\ref{eq:cov-derits-spinors}). In the standard Einstein-Cartan theory the term $T$ does not appear in $D\psi$. But in this work we especially pay attention in order to include all bases $\{I, \sigma_{ab}\}$ of $Cl^+(1,2)$ that are the generators of $Spin_+(1,2)$ group doubly covering the restricted special Lorentz group, $SO_+(1,2)$. In fact, since in the Clifford algebra $Cl^+(1,2)$ there is more structure than in the matrix algebra $\text{Mat}(2,\mathbb{R})$, it is easier to catch the term $T$ in it rather than the matrix notation. Nevertheless, since the connection carries effect of gravitational field, all possible interactions between gravity and spinor field are taken into account in the formula (\ref{eq:cov-derits-spinors}). This extra contribution in the definition of covariant exterior derivative of a spinor is a novel modification.  For more discussions on extended covariant derivative of a spinor, one can consult \cite{koivisto-jimenez2020}. As a final remark we calculate the curvature of spinor bundle
 \begin{align}
     \mathbb{D}^2\psi= \frac{1}{2} \left( R^{ab} \sigma_{ab} - dT \right)\psi .
 \end{align}

\section{Scale invariant Einstein-Cartan theory}

In this section we firstly introduce a scalar field $\phi(x)$ transforming as $\phi' = W^{-1} \phi$ under a $W(2,2)$-transformation. Then its $W(2,2)$-covariant exterior derivative becomes,
 \begin{align}
     \mathbb{D}\phi = d\phi - \frac{1}{2} T \phi, \label{eq:cov-deriv-scalar}
 \end{align}
such that $\mathbb{D}\phi' = W^{-1} (\mathbb{D}\phi)$. Now we formulate the scale invariant Einstein-Cartan theory by the Lagrangian 3-form by combining minimally the Einstein-Hilbert Lagrangian, the Dirac Lagrangian and the scalar field Lagrangian,
 \begin{align}
     L = L_{EH} + L_D + L_\phi \label{eq:lagrange-nonrieman}
 \end{align}
where
 \begin{subequations}
    \begin{align}
        L_{EH} &= -\frac{1}{2\kappa} \phi R^a{}_b \wedge *e_a{}^b, \label{eq:eins-hilbert1} \\
        L_D &= \frac{i}{2} \left[ \overline{\psi} *\gamma \wedge \left(\mathbb{D}\psi \right) -  \left(\mathbb{D}\overline{\psi} \right) \wedge *\gamma \psi \right] + im \phi \overline{\psi} \psi *1, \label{eq:dirac-lag1} \\
        L_\phi &= \phi^{-1} \mathbb{D}\phi \wedge *\mathbb{D}\phi + \mu \phi^3 *1 .
    \end{align}
 \end{subequations}
Here $\kappa$ is gravitational coupling constant, $m$ is mass of spinor field, $\mu$ is a constant that can be interpreted as mass of the scalar field and $\gamma := \gamma_a e^a$ is $Cl(1,2)$-valued 1-form. We introduce imaginary unit in $L_D$ in order to make it Hermitian, $L_D^\dagger = L_D$. We perform independent variations of $L$ with respect to $e^a$, $\omega^{ab}$, $\overline{\psi}$ and $\phi$. Then by using $\delta L=0$ and by discarding exact forms we obtain field equations, 
  \begin{subequations}
  \begin{align}
      -\frac{1}{2\kappa} \epsilon_{abc} \phi R^{bc} + \tau_a[\psi] + \tau_a[\phi] &=0, & &\text{COFRAME} \label{eq:first-eqn} \\
      -\frac{1}{2\kappa} \left( \epsilon_{abc} \phi T^{c} + d\phi \wedge *e_{ab} \right) + \Sigma_{ab}[\psi] + \Sigma_{ab}[\phi] &=0, & &\text{CONNECTION} \label{eq:second-eqn} \\
      i*\gamma \wedge  \mathbb{D} \psi + i m \phi\psi *1 &=0, & &\text{DIRAC} \label{eq:dirac-eqn} \\
      -\frac{1}{2\kappa} R^{ab} \wedge *e_{ab} + im \overline{\psi} \psi *1 + \nabla[\phi] &= 0, & &\text{SCALAR} \label{eq:scalar-eqn}
  \end{align}
  \end{subequations}
where energy-momentum 2-forms, $\tau_a[\psi]:=\partial L_D/\partial e^a$, $\tau_a[\phi]:=\partial L_\phi/\partial e^a$, and angular momentum 2-forms, $\Sigma_{ab}[\psi]:= \partial L_D / \partial \omega^{ab}$, $\Sigma_{ab}[\phi]:= \partial L_\phi / \partial \omega^{ab}$, for spinor and scalar fields, and the 3-form of scalar field, $\Delta[\phi] := \partial L_\phi / \partial \phi$, are obtained, respectively,
  \begin{subequations}
    \begin{align}
        \tau_a[\psi] :=& \frac{i}{2} \left[ \overline{\psi} \gamma^b \left(\mathbb{D}\psi\right) - \left(\mathbb{D}\overline{\psi}\right) \gamma^b \psi \right] \wedge *e_{ab} + im \phi \rho *e_a  \\
         \tau_a[\phi] :=&  - \phi^{-1} \left[ (\iota_a\mathbb{D}\phi) \wedge *\mathbb{D}\phi + \mathbb{D}\phi \wedge (\iota_a*\mathbb{D}\phi) \right] + \mu \phi^3 *e_a \nonumber \\
          &+  \mathbb{D}*(\mathbb{D}\phi \wedge e_a) - (\iota_a T) \wedge *\mathbb{D}\phi - (\iota_aT^b) \wedge \iota_b *\mathbb{D}\phi \label{eq:ner-momon-scalar} \\
        \Sigma_{ab}[\psi] :=& -\frac{i}{4} \rho e_{ab} ,\\
         \Sigma_{ab}[\phi] :=& \frac{1}{2} \left[ e_b \wedge *(\mathbb{D}\phi \wedge e_a) - e_a \wedge *(\mathbb{D}\phi \wedge e_b) \right] , \\
         \Delta[\phi] :=& -\phi^{-2} \mathbb{D}\phi \wedge *\mathbb{D}\phi -2 d (\phi^{-1} *\mathbb{D}\phi) - T \phi^{-1} *\mathbb{D}\phi + 3\mu \phi^2 *1.
    \end{align}
  \end{subequations}
Here it is worthy to remark that the term $T/2$ does not appear in the variational Dirac equation (\ref{eq:dirac-eqn}). This is a resolution of the inconsistency problem of Einstein-Cartan theory stated in Introduction. Furthermore, it is worthwhile to remark that the presence of torsion trace 1-form in (\ref{eq:cov-deriv-scalar}) manifests the non-minimal coupling. Otherwise, angular momentum (or spin current) 2-form of scalar field would have been zero.

Now, torsion is solved analytically by some algebra in terms of spinor and scalar fields from CONNECTION equation (\ref{eq:second-eqn}),  
  \begin{align}
      T^a = \frac{i\kappa}{2} \phi^{-1} \rho *e^a - \phi^{-1} d\phi \wedge  e^{a} . \label{eq:torsion2}
  \end{align}
In general torsion 2-form can be split to three pieces
 \begin{align}
     T^a = \overset{(1)}{T^a} + \overset{(2)}{T^a} + \overset{(3)}{T^a} .
 \end{align}
In three dimensions they are written as
 \begin{align}
     \overset{(2)}{T^a} = - \frac{1}{2} T \wedge e^a , \qquad \overset{(3)}{T^a} =  \frac{1}{3} \iota^a \mathcal{T}  , \qquad  \overset{(1)}{T^a} = T^a - \overset{(2)}{T^a} - \overset{(3)}{T^a} ,
 \end{align}
where $T :=\iota_a T^a$ trace 1-form and $\mathcal{T} := e_a \wedge T^a$ trace 3-form. Since $T$ has three components and $\mathcal{T}$ has only one component, $\overset{(2)}{T^a}$ is vector piece and $\overset{(3)}{T^a}$ is scalar piece (the so-called axial vector component in four-dimensions), respectively. $\overset{(1)}{T^a}$ with five components is tensor piece. Correspondingly, our torsion (\ref{eq:torsion2}) has the second and the third pieces and no the first piece because of 
  \begin{align}
   T = 2 \phi^{-1} d\phi \qquad \text{and} \qquad \mathcal{T} = \frac{3i}{2}\kappa \phi^{-1}\rho *1 .
      \label{eq:trace-torsion}
  \end{align}
When one compares these results with (27) of Ref.\cite{dereli-ozdemir-2013}, it observed that the non-vanishing trace 1-form  along with trace 3-form of torsion is a new outcome. Substitution of the result (\ref{eq:torsion2}) to (\ref{eq:cov-deriv-scalar}) yields $\mathbb{D}\phi=0$ under which other field equations turn out to be simpler forms, 
    \begin{subequations}
  \begin{align}
      -\frac{1}{2\kappa} \epsilon_{abc} \phi R^{bc} + \tau_a[\psi] + \mu \phi^3 *e_a &=0, & &\text{COFRAME} \label{eq:first-eqn2i} \\
            i*\gamma \wedge  \mathbb{D} \psi + i m \phi\psi *1 &=0, & &\text{DIRAC} \label{eq:dirac-eqn2i} \\
      -\frac{1}{2\kappa} R^{ab} \wedge *e_{ab} + im \rho *1 + 3\mu \phi^2 *1 &= 0. & &\text{SCALAR} \label{eq:scalar-eqn2i}
  \end{align}
  \end{subequations}
From the $W(2,2)$-gauge symmetry point of view, in an truly scale-invariant model it is expected that one can always gauge away the scalar field $ \phi$ and bring it to a constant value, $\phi = \phi_0$, with the help of an appropriate scale parameter $W$. This means that the number of independent field equations is actually less than the number of original variables. Correspondingly, we have checked that the SCALAR equation (\ref{eq:scalar-eqn2i}) is just a trace part of COFRAME equation (\ref{eq:first-eqn2i}).

\section{Riemannian formulation of the theory}

Firstly we calculate contortion 1-form by substituting (\ref{eq:torsion2}) to (\ref{eq:contortion})
 \begin{align}
     K_{ab} = \phi^{-1} \left[  -\frac{i}{4} \kappa \rho *e_{ab} -  (\partial_a \phi) e_b + (\partial_b\phi) e_a \right]  \label{eq:contort2}
 \end{align}
where $\partial_a\phi := \iota_a d\phi$. Then by noticing $\widetilde{D}e^a=0$, $\widetilde{D}*e^a=0$,  $\widetilde{D}*e^{ab}=0$ and $d\phi \wedge *e_{ab}= (\partial_b \phi) *e_a - (\partial_a\phi) *e_b$ the related quantities are computed 
  \begin{subequations}
  \begin{align}
      R_{ab} =& \widetilde{R}_{ab} + \frac{i\kappa}{4} (2\rho \phi^{-2}  d\phi -\phi^{-1} d\rho  ) \wedge *e_{ab} + \phi^{-1} \left[  \widetilde{D}\left(\partial_b \phi\right) \wedge e_a - \widetilde{D} \left(\partial_a \phi\right) \wedge e_b \right]  \nonumber \\
     &- \frac{\kappa^2}{16}  \rho^2 \phi^{-2} e_{ab} + 2 \phi^{-2} d\phi \wedge \left[ (\partial_a\phi) e_b - (\partial_b \phi) e_a \right] - \phi^{-2} (\partial \phi)^2 e_{ab} . \label{eq:decomp-curva2} \\
      \mathbb{D} \psi =& \widetilde{D}\psi + \frac{i}{8} \kappa \rho \phi^{-1} \gamma_a \psi e^a + \frac{1}{2} \phi^{-1} (\partial^a \phi) \gamma^b \psi *e_{ab} -  \phi^{-1} d\phi \psi, \label{eq:Dpsi-decomp}\\
       \mathbb{D} \overline{\psi} =& \widetilde{D}\overline{\psi} - \frac{i}{8} \kappa \rho \phi^{-1} \overline{\psi} \gamma_a e^a - \frac{1}{2} \phi^{-1} (\partial^a \phi) \overline{\psi} \gamma^b   *e_{ab} - \overline{\psi} \phi^{-1} d\phi , \label{eq:Dpsi-bar-decomp} \\
       \tau_a[\psi] =& \widetilde{\tau}_a[\psi] - \frac{1}{4}\kappa  \rho^2 \phi^{-1} *e_a - \frac{i}{2} \rho \phi^{-1} d\phi \wedge e_a ,
  \end{align}
 \end{subequations}
where $\widetilde{D} \left(\partial_a \phi\right) := d \left(\partial_a \phi\right) - \widetilde{\omega}^c{}_a \left(\partial_c \phi\right)$ and $(\partial \phi)^2 := (\partial_c\phi)(\partial^c\phi)$ and
  \begin{subequations}
   \begin{align}
        \widetilde{D}\psi &:= d\psi + \frac{1}{2} \widetilde{\omega}^{ab} \sigma_{ab} \psi \qquad \text{and} \qquad  \widetilde{D}\overline{\psi} := d\overline{\psi} - \frac{1}{2}  \overline{\psi}\sigma_{ab} \widetilde{\omega}^{ab}, \\
         \widetilde{\tau}_a[\psi] &:= \frac{i}{2} \left[ \overline{\psi} \gamma^b \left(\widetilde{D}\psi\right) - \left(\widetilde{D}\overline{\psi}\right) \gamma^b \psi \right] \wedge *e_{ab} + im  \rho \phi *e_a . \label{eq:energy-moment-spinor-riemann}
   \end{align}
    \end{subequations}
We insert these results into (\ref{eq:first-eqn2i}), and by rearranging terms we find the decomposed COFRAME equation,
   \begin{align}
    \widetilde{R}_{ab} =& -\kappa \phi^{-1} \epsilon_{abc}  \widetilde{\tau}^c[\psi] - \frac{3}{16} \kappa^2  \rho^2 \phi^{-2} e_{ab}   + \frac{i\kappa}{4}  \phi^{-1} d\rho \wedge *e_{ab} + \phi^{-1} \left[ \widetilde{D} \left(\partial_a \phi\right) \wedge e_b  -\widetilde{D}\left(\partial_b \phi\right) \wedge e_a \right] \nonumber \\
    &+\phi^{-2}\{ (\partial \phi)^2 e_{ab} - 2 d\phi \wedge \left[ (\partial_a\phi) e_b - (\partial_b \phi) e_a \right] \} +  \mu \kappa \phi^2 e_{ab} . \label{eq:decomp-coframe2i}
 \end{align}
Now we decompose DIRAC equation (\ref{eq:dirac-eqn2i}) by using (\ref{eq:Dpsi-decomp}),
 \begin{align}
     i*\gamma \wedge \widetilde{D}\psi +  im\phi \psi *1 - \frac{3}{8}\kappa \rho \phi^{-1} \psi *1 =0. \label{eq:dirac-decompsi}
 \end{align}
Finally we decompose SCALAR equation (\ref{eq:scalar-eqn2i}),
 \begin{align}
     -\frac{1}{2\kappa} \widetilde{R}^{ab} \wedge *e_{ab} + \frac{3}{16}\kappa \rho^2  \phi^{-2} *1 + im\rho *1  - \frac{1}{\kappa} \phi^{-2} d\phi \wedge *d\phi +\frac{2}{\kappa} \phi^{-1} d*d\phi + 3\mu \phi^2 *1 =0 \label{eq:scalar-decomps}
 \end{align}

At this stage the decomposition of the Lagrangian (\ref{eq:lagrange-nonrieman}) is calculated up to a closed form,
 \begin{align}
   \widetilde{L} = \widetilde{L}_{EH} + \widetilde{L}_D - \frac{3}{16}\kappa  \rho^2 \phi^{-1} *1   - \frac{1}{\kappa} \phi^{-1} d\phi \wedge *d\phi + \mu \phi^3 *1 + \lambda_a \wedge T^a , \label{eq:lagrang-riemann}
 \end{align}
where $\lambda_a$ is a Lagrange multiplier 1-form constraining zero-torsion, and the Riemannian Einstein-Hilbert Lagrangian and the Dirac Lagrangian are, respectively,
 \begin{align}
     \widetilde{L}_{EH} &= -\frac{1}{2\kappa} \phi \widetilde{R}^{ab} \wedge *e_{ab} , \\
      \widetilde{L}_D &= \frac{i}{2} \left[ \overline{\psi} *\gamma \wedge \left(\widetilde{D}\psi \right) -  \left(\widetilde{D}\overline{\psi} \right) \wedge *\gamma \psi \right] + im \phi \overline{\psi} \psi *1. \label{eq:dirac-lag-rieman}
 \end{align}
The third and fourth terms in (\ref{eq:lagrang-riemann}) represent the the existence of torsion. One can follow the torsional effects by tracing these terms in the Riemannian spacetime geometry. $\lambda_a$-variation of $\widetilde{L}$ warrants that the connection is Levi-Civita, $\widetilde{\omega}^a{}_b$. Then, $\overline{\psi}$-variation and $\phi$-variation yield DIRAC equation (\ref{eq:dirac-decompsi}) and SCALAR equation (\ref{eq:scalar-decomps}), respectively. Thus $e^a$ and $\omega^{ab}$ variations causes to following equations, respectively, 
    \begin{subequations}
  \begin{align}
        -\frac{\phi}{2\kappa} \epsilon_{abc}\widetilde{R}^{bc} + \widetilde{\tau}_a[\psi] - \frac{3}{16} \kappa \rho^2 \phi^{-1} *e_a + \frac{\phi^{-1}}{\kappa} \widetilde{\tau}_a[\phi] + \mu \phi^3 *e_a + \widetilde{D}\lambda_a  &=0, \label{eq:coframe3-eqn} \\
      - \frac{1}{\kappa} d\phi \wedge *e_{ab} -\frac{i}{2} \rho e_{ab} + e_b \wedge \lambda_a - e_a \wedge \lambda_b &=0, \label{eq:connection3-eqn} 
  \end{align}
  \end{subequations}
where
 \begin{align}
        \widetilde{\tau}_a[\phi] := \iota_ad\phi \wedge *d\phi + d\phi \wedge \iota_a *d\phi = 2(\partial_a\phi) (\partial_b \phi) *e^b -(\partial\phi)^2 *e_a .
 \end{align}
The Lagrange multiplier could be computed from the second equation (\ref{eq:connection3-eqn}) by hitting $\iota_{ab}$,
 \begin{align}
     \lambda_a = -\frac{i}{4} \rho e_a + \frac{1}{\kappa} (\partial^b \phi) *e_{ab} .
 \end{align}
Then it turns out to be
 \begin{align}
     \widetilde{D}\lambda_a = - \frac{i}{4} d\rho \wedge e_a + \frac{1}{\kappa} \widetilde{D}(\partial^b \phi) \wedge *e_{ab} .
 \end{align}
Finally the usage of this result in the equation (\ref{eq:coframe3-eqn}) and rearrangement of the terms yield  the equation (\ref{eq:decomp-coframe2i}) as expected. Consequently, we studied the same theory in two different geometries and see that they are equivalent in both the Lagrangian level and the equation level. Meanwhile it is observed that the non-Riemannian formalism (\ref{eq:lagrange-nonrieman}) looks tidier than the Riemannian one (\ref{eq:lagrang-riemann}).

\section{Discussion}

Since the general relativity does not predict gravitational wave in empty space in three dimensions, the three-dimensional extended gravity models attract pretty much attention. Therefore we treated the Einstein-Cartan theory in three dimensions by starting with a discussion about symmetry group. Then we concluded that the complete gauge group should include the scale group along with the Lorentz group. It is the Weyl group, $W(2,2)=SO_+(1,2) \otimes \mathbb{R}^+$ with four parameters. At this point we postulated scale transformation of the affine connection 1-form so as to leave the metricity condition invariant. We also extended the covariant exterior derivative of spinor by adding the term, $-T\psi/2$, where $T=\iota_a T^a$ is the torsion trace 1-form. Thus, we saw that our new definition of $\mathbb{D}\psi$ given by the equation (\ref{eq:cov-derits-spinors}) resolves the inconsistency problem in the Einstein-Cartan theory which can be stated that Dirac equations obtained from the equation level and the Lagrangian level are not same. Afterwards, we wrote a $W(2,2)$-invariant Lagrangian by introducing a compensating scalar field, $\phi$. We computed the field equations by independent variations and could solve torsion algebraically. Substitution of this result into other equations simplified them significantly. In the subsequent section, we decomposed all concerned non-Riemannian quantities as Riemannian quantity plus torsional contribution. Accordingly, we rewrote COFRAME, DIRAC and SCALAR equations, and also the Lagrangian 3-form in the Riemannian geometry with novel contributions. Finally we verified that the decomposed field equations are the variational field equations of the decomposed Lagrangian. Consequently, we showed the equivalence of two formulations of the same theory. Of course, the non-Riemannian formulation seems much tidier, but one can gain physical insights about torsion tensor by tracing novel terms in the Riemannian formulation.

In our consideration trace 1-form, $T$, of torsion behaves like a gauge potential of scale transformation. Meanwhile, it is known from application procedure of gauge theory that one should add kinetic counterpart, $dT$, of gauge potential to Lagrangian. That is, a scale invariant term, $ \frac{\nu}{2} \phi^{-1} dT \wedge *dT$, is expected to be in the Lagrangian (\ref{eq:lagrange-nonrieman}) where $\nu$ is a coupling constant. But when we add that term, the connection variation yields following extra term to CONNECTION equation (\ref{eq:second-eqn})
 \begin{align}
     \frac{\nu}{2} \left[ e_a \wedge \iota_b d(\phi^{-1} *dT) - e_b \wedge \iota_a d(\phi^{-1} *dT) \right] .
 \end{align}
Thus since now torsion gains propagating degrees of freedom because of the contributions $dT$ and $d*dT$, one can not solve it algebraically anymore. Generalisation of our model with inclusion of this term and some explicit solutions are left as our future project.

\section*{Appendix}

 \appendix

\section{Clifford algebra, Lorentz group, spin group}

Let $\left\{ 1, \gamma_0, \gamma_1, \gamma_2 \right\}$ be generators of the eight dimensional Clifford algebra, $Cl(1,2)$, such that
 \begin{align}
     \gamma_a \gamma_b + \gamma_b \gamma_a = 2 \eta_{ab} I, \qquad a,b = 0,1,2,
 \end{align}
where $\eta_{ab}$ is the components of Minkowski metric, $\eta_{00}=-1$, $\eta_{11}=+1$, $\gamma_{22}=+1$, $\text{others}=0$. At the beginning the gamma objects need not to be some matrices, they are abstract objects in general. A basis for $Cl(1,2)$ is given by the set
 \begin{align}
     \left\{ 1, \gamma_0, \gamma_1, \gamma_2, \gamma_{01}, \gamma_{02}, \gamma_{12}, \gamma_{012} \right\}
 \end{align}
where $1$ is the basis for scalars, $\gamma_a$ are the bases for 1-vectors, $\gamma_{ab}:= \gamma_a\gamma_b$ ($a \neq b$) are the bases for bivectors and $\gamma_{012}:= \gamma_0 \gamma_1 \gamma_2$ is the basis for 3-vectors \cite{lounesto1997}.   Its multiplication table is below.
  \begin{center}
 \begin{tabular}{ c|cccccccc } 
 & $1$  & $\gamma_0$ & $\gamma_1$ & $\gamma_2$ & $\gamma_{01}$ & $\gamma_{02}$ & $\gamma_{12}$ & $\gamma_{012}$ \\ 
 \hline
 $1$ & $1$  & $\gamma_0$ & $\gamma_1$ & $\gamma_2$ & $\gamma_{01}$ & $\gamma_{02}$ & $\gamma_{12}$ & $\gamma_{012}$ \\
  $\gamma_0$ & $\gamma_0$  & $-1$ & $\gamma_{01}$ & $\gamma_{02}$ & $-\gamma_{1}$ & $-\gamma_{2}$ & $\gamma_{012}$ & $-\gamma_{12}$ \\
   $\gamma_1$ & $\gamma_1$  & $-\gamma_{01}$ & $1$ & $\gamma_{12}$ & $-\gamma_{0}$ & $-\gamma_{012}$ & $\gamma_{2}$ & $-\gamma_{02}$ \\
 $\gamma_2$ & $\gamma_2$  & $-\gamma_{02}$ & $-\gamma_{12}$ & $1$ & $\gamma_{012}$ & $-\gamma_{0}$ & $-\gamma_{1}$ & $\gamma_{01}$ \\
 $\gamma_{01}$ & $\gamma_{01}$  & $\gamma_{1}$ & $\gamma_{0}$ & $\gamma_{012}$ & $1$ & $\gamma_{12}$ & $\gamma_{02}$ & $\gamma_{2}$ \\
  $\gamma_{02}$ & $\gamma_{02}$  & $\gamma_{2}$ & $-\gamma_{012}$ & $\gamma_{0}$ & $-\gamma_{12}$ & $1$ & $-\gamma_{01}$ & $-\gamma_{1}$ \\
  $\gamma_{12}$ & $\gamma_{12}$  & $\gamma_{012}$ & $-\gamma_{2}$ & $\gamma_{1}$ & $-\gamma_{02}$ & $\gamma_{01}$ & $-1$ & $-\gamma_{0}$ \\
  $\gamma_{012}$ & $\gamma_{012}$  & $-\gamma_{12}$ & $-\gamma_{02}$ & $\gamma_{01}$ & $\gamma_{2}$ & $-\gamma_{1}$ & $-\gamma_{0}$ & $1$ \\
 \end{tabular}
 \end{center}
Thus $Cl(1,2)$ is also a group, the so-called $pin(1,2)$ group, as well as a eight-dimensional real linear vector space. In particle physics community it is a custom to denote $\gamma_{ab} := 2\sigma_{ab}$ and $\gamma_{012} := \gamma_5$. Now we consider the even subset, i.e., 0-vectors and 2-vectors, $\left\{ 1, \gamma_{01}, \gamma_{02}, \gamma_{12} \right\}$ and look at its multiplication table.
  
   \begin{center}
 \begin{tabular}{ c|cccc } 
    & $1$ & $ \gamma_{01} $ & $ \gamma_{02} $ & $ \gamma_{12} $ \\
    \hline
  $1$ & $1$ & $ \gamma_{01} $ & $ \gamma_{02} $ & $ \gamma_{12} $ \\
  $ \gamma_{01}$  & $ \gamma_{01}$  & $ 1 $ & $ \gamma_{12} $ & $ \gamma_{02}$ \\
  $ \gamma_{02}$  & $ \gamma_{02}$  & $ -\gamma_{12} $ & $ 1$ & $ -\gamma_{01}$ \\
  $ \gamma_{12}$  & $ \gamma_{12}$  & $ -\gamma_{02} $ & $ \gamma_{01} $ & $-1$ \\
    \end{tabular}
   \end{center}
Similarly, this subset also forms an four-dimensional subalgebra, $Cl^+(1,2)$, and a subgroup, $pin_+(1,2)$. A general element of $Cl^+(1,2)$ can be written as
  \begin{align}
      u= u_0 + u_{01} \gamma_{01} + u_{02} \gamma_{02} + u_{12} \gamma_{12}
  \end{align}
where the scalars, $\left\{ u_0, u_{01}, u_{02}, u_{12}\right\}$, are components of $u$. That is a direct sum of a scalar $u_0$ and a bivector, $u_{01} \gamma_{01} + u_{02} \gamma_{02} + u_{12} \gamma_{12}$. One of useful operations on $u$ is its reverse defined by
  \begin{align}
      \widehat{u} = u_0 - u_{01} \gamma_{01} - u_{02} \gamma_{02} - u_{12} \gamma_{12} .
  \end{align}
Besides, $Cl^+(1,2)$ is isomorphic to $2\times 2$-real matrices, $Cl^+(1,2) \simeq \text{Mat}(2,\mathbb{R})$.
One set of representation is as follows
 \begin{align}
  1 \simeq 
     \begin{bmatrix}
       1 & 0 \\
       0 & 1
     \end{bmatrix}, \qquad 
     \gamma_0 \simeq 
     \begin{bmatrix}
       0 & 1 \\
       -1 & 0
     \end{bmatrix}, \qquad
     \gamma_1 \simeq 
     \begin{bmatrix}
       0 & 1 \\
       1 & 0
     \end{bmatrix}, \qquad
     \gamma_2 \simeq 
     \begin{bmatrix}
       1 & 0 \\
       0 & -1
     \end{bmatrix}. \label{eq:dirac-matricies}
 \end{align}
Then abstract $u \in Cl^+(1,2)$ may be written by a matrix
  \begin{align}
      u \simeq \begin{bmatrix}
       u_0 + u_{01} & -u_{02} - u_{12} \\
       -u_{02} + u_{12} & u_{0} - u_{01}
     \end{bmatrix} = 
      \begin{bmatrix}
       a & b \\
       c & d
     \end{bmatrix}.
  \end{align}
In order to see the meaning of reverse operation of $Cl^+(1,2)$ in matrix formalism we redefined the components,
 \begin{align}
     a= u_0 + u_{01}, \qquad b= -u_{02} - u_{12} , \qquad c= -u_{02} + u_{12} , \qquad d= u_{0} - u_{01} .
 \end{align}
Thus, the reverse of $u$ takes the form
 \begin{align}
     \widehat{u} \simeq \begin{bmatrix}
       u_0 - u_{01} & u_{02} + u_{12} \\
       u_{02} - u_{12} & u_{0} + u_{01}
     \end{bmatrix} = 
      \begin{bmatrix}
       d & -b \\
       -c & a
     \end{bmatrix} = (ad-bc) \begin{bmatrix}
       a & b \\
       c & d
     \end{bmatrix}^{-1} \quad \text{for} \quad ad-bc \neq 0.
 \end{align}
In this case the condition $\widehat{u} u = u \widehat{u} =1$ corresponds to $\text{det}[u]=+1$ in matrix notation.

A restricted or special orthochronous Lorentz group, $SO_+(1,2)$, which preserves both time and space orientations is defined by
 $$
   SO_+(1,2) = \left\{ L \in \text{Mat}(3, \mathbb{R}) \; \big| \;  L^T \eta L = \eta , \quad \text{det} L = +1 \right\}.
 $$
The spacetime vector $\mathbf{x} \in \mathbb{R}^{1,2}$ transforms according to $\mathbf{x}' = L \mathbf{x}$. Any $L \in SO_+(1,2)$ can be written as an exponential $L=e^A$ of a Minkowski-antisymmetric matrix\footnote{For the notation $S=e^{\frac{1}{2}\sigma_{ab}\vartheta^{ab}}$ used in the section of Introduction there are correspondences $\vartheta_{01} = a_1$, $\vartheta_{02} = a_2$, $\vartheta_{12} = b$.}
 \begin{align}
   A=
     \begin{bmatrix}
     0 & a_1 & a_2 \\
     a_1 & 0 & -b \\
     a_2 & b & 0
     \end{bmatrix} \quad \text{satisfying} \quad \eta A^T \eta^{-1} = -A.
 \end{align}
The matrix $A$ could be specified by a vector $\Vec{a}=a_1 \gamma_1 + a_2 \gamma_2 \in \mathbb{R}^2$ and a scalar $b \in \mathbb{R}$. If $b=0$, then $L$ is a boost at velocity $\left| \Vec{v} \right| = c \tanh\left| \Vec{a}\right|$. If $\Vec{a}=0$, then $L \in SO(2)$ is a rotation of the Euclidean space $\mathbb{R}^2$ by the angle $b$.  For example, $x$-boost, $\mathbf{x}' = L_{a_x} \mathbf{x}$ could be calculated explicitly by
  \begin{align}
      \begin{bmatrix}
       ct' \\
       x' \\
       y'
      \end{bmatrix} = \begin{bmatrix}
       \Gamma & -\beta_x \Gamma & 0 \\
       -\beta_x \Gamma & \Gamma & 0 \\
       0 & 0 & 1
      \end{bmatrix} \begin{bmatrix}
       ct \\
       x \\
       y
      \end{bmatrix}
    \end{align}
where $\Gamma := 1/\sqrt{1-\beta_x^2}$, $\beta_x := v_x/c$ and $\tanh{a_x} = \beta_x$ giving $\cosh{a_x}:=\Gamma$ and $\sinh{a_x}:=\Gamma \beta_x$. Similarly, rotation of $\mathbf{x}$ from $x$ to $y$ by angle $b$ formulated as $\mathbf{x}' = L_{b} \mathbf{x}$ is computed by
  \begin{align}
      \begin{bmatrix}
       ct' \\
       x' \\
       y'
      \end{bmatrix} = \begin{bmatrix}
       1 & 0 & 0 \\
       0 & \cos{b} & -\sin{b} \\
       0 & \sin{b} & \cos{b}
      \end{bmatrix} \begin{bmatrix}
       ct \\
       x \\
       y
      \end{bmatrix}.
    \end{align}

The Lorentz group, $SO_+(1,2)$, has a double cover
 $$
   Spin_+(1,2) = \left\{ S \in Cl^+(1,2) \; \big| \; S \widehat{S}=1 \right\} .
 $$
Under a $SO_+(1,2)$ Lorentz transformation induced by $S \in Spin_+(1,2)$ the spacetime vector $\mathbf{x}$ transforms in accordance with $\mathbf{x}' = S \mathbf{x} S^{-1}$. As a boost at velocity, $\Vec{v} \in \mathbb{R}^2$, can be computed by $S=e^{\frac{1}{2}\Vec{a}\gamma_0}$, where $\Vec{a}=\arctanh\left(\frac{\Vec{v}}{c}\right)$, a rotation of amount $b$ by $S=e^{\frac{1}{2}b\gamma_{12}}$. For example, $x$-boost, $\mathbf{x}' = S_{a_x} \mathbf{x} S_{a_x}^{-1}$, can be expressed explicitly,
 \begin{subequations}
 \begin{align}
   ct' \gamma_0 + x' \gamma_1 + y' \gamma_2 = e^{\frac{1}{2}a_x \gamma_{10}} \left( ct \gamma_0 + x \gamma_1 + y \gamma_2 \right) e^{-\frac{1}{2}a_x \gamma_{10}} , \\
   \begin{bmatrix}
   y' & ct' + x' \\
   -ct' + x' & -y'
   \end{bmatrix} =   
    S_{a_x}
   \begin{bmatrix}
   y & ct + x \\
   -ct + x & -y
   \end{bmatrix} S_{a_x}^{-1},
 \end{align}
 \end{subequations}
where 
  \begin{subequations}
 \begin{align}
     S_{a_x} &= e^{\frac{1}{2}a_x \gamma_{10}} = \begin{bmatrix}
    \cosh{\frac{a_x}{2}} - \sinh{\frac{a_x}{2}} & 0 \\
    0 &  \cosh{\frac{a_x}{2}} + \sinh{\frac{a_x}{2}} 
   \end{bmatrix} , \\
   S_{a_x}^{-1} &= e^{-\frac{1}{2}a_x \gamma_{10}} = \begin{bmatrix}
    \cosh{\frac{a_x}{2}} + \sinh{\frac{a_x}{2}} & 0 \\
    0 &  \cosh{\frac{a_x}{2}} - \sinh{\frac{a_x}{2}} 
   \end{bmatrix} .
 \end{align}
  \end{subequations}
Together with $\tanh{a_x} := \beta_x = {v_x}/{c}$ meaning $\cosh{a_x} := \Gamma = 1/ \sqrt{1-v_x^2/c^2}$ and $\sinh{a_x} := \Gamma \beta_x$, matrix multiplications yield the $x$-boost equations in the very well known form. One may check $\text{det}[S] = +1$ in the matrix algebra corresponding to $\widehat{S} S =1$ in the Clifford algebra. Similarly, a rotation of $\mathbf{x}$ in $xy$-plane by angle $b$ from $\gamma_1$ to $\gamma_2$, $\mathbf{x}' = S_{b} \mathbf{x} S_{b}^{-1}$, is computed by 
  %\begin{subequations}
 \begin{align}
     S_b = e^{\frac{1}{2}b \gamma_{21}} = \begin{bmatrix}
    \cos{\frac{b}{2}}  &  \sin{\frac{b}{2}} \\
     -\sin{\frac{b}{2}} &  \cos{\frac{b}{2}}  
   \end{bmatrix} , \qquad
   S_b^{-1} = e^{-\frac{1}{2}b \gamma_{21}} = \begin{bmatrix}
    \cos{\frac{b}{2}}  &  -\sin{\frac{b}{2}} \\
     \sin{\frac{b}{2}} &  \cos{\frac{b}{2}}
   \end{bmatrix} .
 \end{align}
 % \end{subequations}

\section*{Acknowledgements}

One of the authors (MA) stays at Istanbul Technical University (ITU) via a sabbatical leave and thanks the Department of Physics Engineering, ITU for warm hospitality. He also thanks Tekin Dereli, Yorgo Senikoglu and Cem Yetismisoglu for enlightening conversations. We thank to the anonymous referee for his/her guiding questions and criticisms.

\end{document}